\documentclass[apj]{emulateapj}

\slugcomment{To Appear in the {\it Astrophysical Journal}}

\shorttitle{Diffuse O VI Emission from the ISM}
\shortauthors{Dixon \& Sankrit}

\newcommand{\fig}[1]{Fig.~\ref{#1}}

\newcommand{\ie}{i.e.}

\newcommand{\ctwo}{\ion{C}{2}}
\newcommand{\ebv}{$E($\bv)}

\newcommand{\flux}{erg cm$^{-2}$ s$^{-1}$  \AA$^{-1}$}
\newcommand{\hone}{\ion{H}{1}}
\newcommand{\htwo}{H$_2$}
\newcommand{\iosix}{$I$(\osix)}
\newcommand{\kms}{km s$^{-1}$}

\newcommand{\nho}{$N$(\hone)}

\newcommand{\oone}{\ion{O}{1}}
\newcommand{\osix}{\ion{O}{6}}
\newcommand{\nosix}{$N$(\osix)}

\newcommand{\sig}{$\sigma$}
\newcommand{\specfit}{{\small SPECFIT}}

\newcommand{\fuse}{{\it FUSE}}

\begin{document}

\title{Recent {\em FUSE}\/ Observations of Diffuse \ion{O}{6}\ Emission
from the Interstellar Medium\altaffilmark{1}}

\author{W. Van Dyke Dixon}
\affil{Department of Physics and Astronomy,
Johns Hopkins University, 3400 N. Charles Street, Baltimore, MD 21218}
\email{wvd@pha.jhu.edu}

\and

\author{Ravi Sankrit\altaffilmark{2}}
\affil{Space Sciences Laboratory, University of California, Berkeley, CA 94720}

\altaffiltext{1}{Based on observations made with the NASA-CNES-CSA {\it Far Ultraviolet Spectroscopic Explorer. FUSE}\/ is operated for NASA by the Johns Hopkins University under NASA contract NAS5-32985.}

\altaffiltext{2}{Visiting Research Scientist, Johns Hopkins University, 3400 N. Charles Street, Baltimore, MD 21218}

\begin{abstract}

We present new results from our survey of diffuse \osix -emitting gas in the interstellar medium with the {\it Far Ultraviolet Spectroscopic Explorer (FUSE).}  Background observations obtained since 2005 have yielded eleven new \osix\ detections of 3\sig\ significance, and archival searches have revealed two more.  An additional 15 sight lines yield interesting upper limits.  Combined with previous results, these observations reveal the large-scale structure of the \osix -bearing gas in the quadrant of the sky centered on the Magellanic Clouds.  The most prominent feature is a layer of low-velocity \osix\ emission extending more than 70\arcdeg\ from the Galactic plane.  At low latitudes ($|b| < 30\arcdeg$), the emission comes from narrow, high-density conductive interfaces in the local ISM.  At high latitudes, the emission is from extended, low-density regions in the Galactic halo.  We also detect \osix\ emission from the interface region of the Magellanic System, a structure recently identified from \hone\ observations.  These are the first detections of emission from high-ionization species in the Magellanic System outside of the Clouds themselves.

\end{abstract}


\section{Introduction}

The lithium-like \osix\ ion is an important tool for studying hot gas in interstellar space, because emission via its 1031.93 and 1037.62 \AA\ resonance lines is the dominant cooling mechanism for gas at temperatures of a few times $10^5$ K \citep{Gnat:Sternberg:07}.  Recent \osix\ absorption-line surveys \citep{Savage:03, Bowen:08} reveal that, within our Galaxy, the \osix -bearing gas has an exponential scale height of 3.2 kpc at negative latitudes and 4.6 kpc at positive latitudes.  The gas is clumpy: spectra of nearly two dozen stars in the Large and Small Magellanic Clouds reveal significant variations in the \osix\ column density of Galactic halo gas over angular scales as small as $3\farcm5$ \citep{Howk:02}.

Closer to the plane, \citet{Bowen:08} find that the average volume density of \osix\ towards stars in the Galactic disk, $n = N($\osix$)/d$, exhibits a dispersion of about 0.26 dex, regardless of the distance to the target stars.  They conclude that the \osix\ does not arise in randomly distributed clouds of a uniform size and density, but is instead distributed among regions with a broad range of column densities, with smaller clouds more common and larger clouds less so.  They envision a turbulent, multiphase medium churned by shock-heated gas from multiple supernova explosions.

\defcitealias{Otte:06}{Paper~I}
\defcitealias{Dixon:06}{Paper~II}

We present the latest results from a survey of diffuse \osix\ emission from the interstellar medium (ISM) conducted with the {\it Far Ultraviolet Spectroscopic Explorer (FUSE)}.\/  \citet[hereinafter Paper~I]{Otte:06} and \citet*[Paper~II]{Dixon:06} discuss observations obtained from launch through 2003 July 25 and 2004 December 31, respectively.  
In these papers, \osix\ emission at 3\sig\ significance is detected along 29 lines of sight.  Measured intensities range from 1.8 to 9.1 kLU, with a median of 3.3 kLU.  An additional 35 sight lines provide upper limits of 2.0 kLU or less.  (One line unit, or photon cm$^{-2}$ s$^{-1}$ sr$^{-1}$, corresponds to $1.9 \times 10^{-11}$ erg cm$^{-2}$ s$^{-1}$ sr$^{-1}$ at 1032 \AA.)  By combining emission and absorption measurements through the same \osix -bearing regions, the authors find evidence for two flavors of \osix -emitting gas: relatively narrow, high-density conductive interfaces in the local ISM and more extended, low-density regions in HVCs.

In this paper, we present data obtained between 2005 January 1 and 2007 July 6, along with two older observations that were missed in the earlier papers.  In \S \ref{sec_observations}, we describe the \fuse\/ instrument, our criteria for selecting sight lines for study, and our data-reduction and analysis techniques.  In \S \ref{sec_discussion}, we derive the physical parameters of the emitting gas and discuss its location along the line of sight.  We present our conclusions in \S \ref{sec_conclusions}.  Because the \fuse\/ pointing system failed in July of 2007, these observations represent the final installment of our survey.

\section{Observations, Data Reduction, and Analysis} \label{sec_observations}

\fuse\/ consists of four coaligned spectrographs. Two employ Al+LiF optical coatings and record spectra over the wavelength range 990--1187 \AA, while the other two use SiC coatings, which provide reflectivity to wavelengths below the Lyman limit. The four channels overlap between 990 and 1070 \AA.  Each channel possesses three apertures that simultaneously sample different parts of the sky. The low-resolution (LWRS) aperture is 30\arcsec\ $\times$ 30\arcsec\ in size. The medium-resolution (MDRS) aperture is 4\arcsec\ $\times$ 20\arcsec\ and lies about $3\farcm5$ from the LWRS aperture. The 1.25\arcsec\ $\times$ 20\arcsec\ high-resolution (HIRS) aperture lies midway between the MDRS and LWRS apertures. A fourth location, the reference point (RFPT), is offset from the HIRS aperture by about 1\arcmin. When a star is placed at the reference point, all three apertures sample the background sky.  For a complete description of \fuse, see \citet{Moos:00} and \citet{Sahnow:00}.

To identify background observations for this study, we searched the Multimission Archive at Space Telescope (MAST) for \fuse\/ data obtained after 2005 January 1 falling in one of three categories: programs S405, S505, S605, S705, and S905, which consist of background observations obtained during spacecraft thermalization periods; MDRS, HIRS, and RFPT observations of point sources, for which the LWRS aperture should sample only background sky; and targets whose names include the strings ``BKG'' or ``SKY.''  To find older observations that may have been missed in our previous studies, we searched the entire archive for target names that include the strings ``OFF,'' ``SKY,'' or ``WIND.''  

We exclude from our sample known supernova remnants, nebulae, and point-source targets in either of the Magellanic Clouds.  Data from programs E124, G052, and G938 will be published elsewhere.  Because its sensitivity at 1032 \AA\ is more than twice that of any other channel, we consider only data from the LiF 1A channel.  We use only data obtained in time-tag mode, which preserves arrival time and pulse-height information for each photon event.  To avoid contamination by geocoronal emission \citep*{Shelton:07}, we use only data obtained during orbital night.  Individual exposures with less than 10 s of night exposure time are excluded from the survey, as are combined data sets with less than 1500 s of night exposure time or with significant continuum flux.  Our final sample consists of data obtained along 120 lines of sight.

The data were reduced using an implementation of CalFUSE v3.2, the latest version of the \fuse\/ data-reduction software package \citep{Dixon:07}, optimized for faint, diffuse emission.  Details are given in \citetalias{Dixon:06}, but we highlight a few points here:  the program operates on one exposure at a time, producing a flux- and wavelength-calibrated spectrum from each.  Background subtraction is turned off.  Zero-point errors in the wavelength scale are corrected by fitting a synthetic emission feature to the Ly $\beta$ airglow line.  The spectra from individual exposures are then shifted to a heliocentric wavelength scale and combined.  

We use an automated routine to search each combined spectrum for statistically significant emission near 1032 \AA.  When we find it, we fit the line with a model emission feature using the non-linear curve-fitting program \specfit\ \citep{Kriss:94}.  The model consists of a top-hat shaped function 106 \kms\ in width, representing the LWRS aperture, convolved with a Gaussian, representing the instrument line-spread function ($\sim$ 25 \kms ) convolved with the intrinsic emission-line profile.  Free parameters in the fit are the level and slope of the background (assumed linear) and the amplitude, central wavelength, and FWHM of the Gaussian.  

While we use the same line identification and fitting algorithms described in \citetalias{Dixon:06}, one change in our technique is worth noting:  Because our count rates are low, we bin the spectra to raise the counts per bin into the regime where Gaussian statistics apply.  For most spectra, we bin by 14 pixels or 53 \kms, a value chosen so that that two binned pixels just span a diffuse emission feature.  If the continuum averages more than 1.5 counts per pixel, we take advantage of the higher signal-to-noise ratio and bin by only 8 pixels.  Because narrow emission features may be undersampled by 14-pixel binning, in \citetalias{Dixon:06} we lowered the threshold for 8-pixel binning to 0.5 counts per bin for spectra containing features with FWHM $< 100$ \kms.  In this paper, we use 14-pixel binning for all faint continua, regardless of the line width.  Implications of this change are discussed below.

\begin{figure}
\epsscale{.85}
\plotone{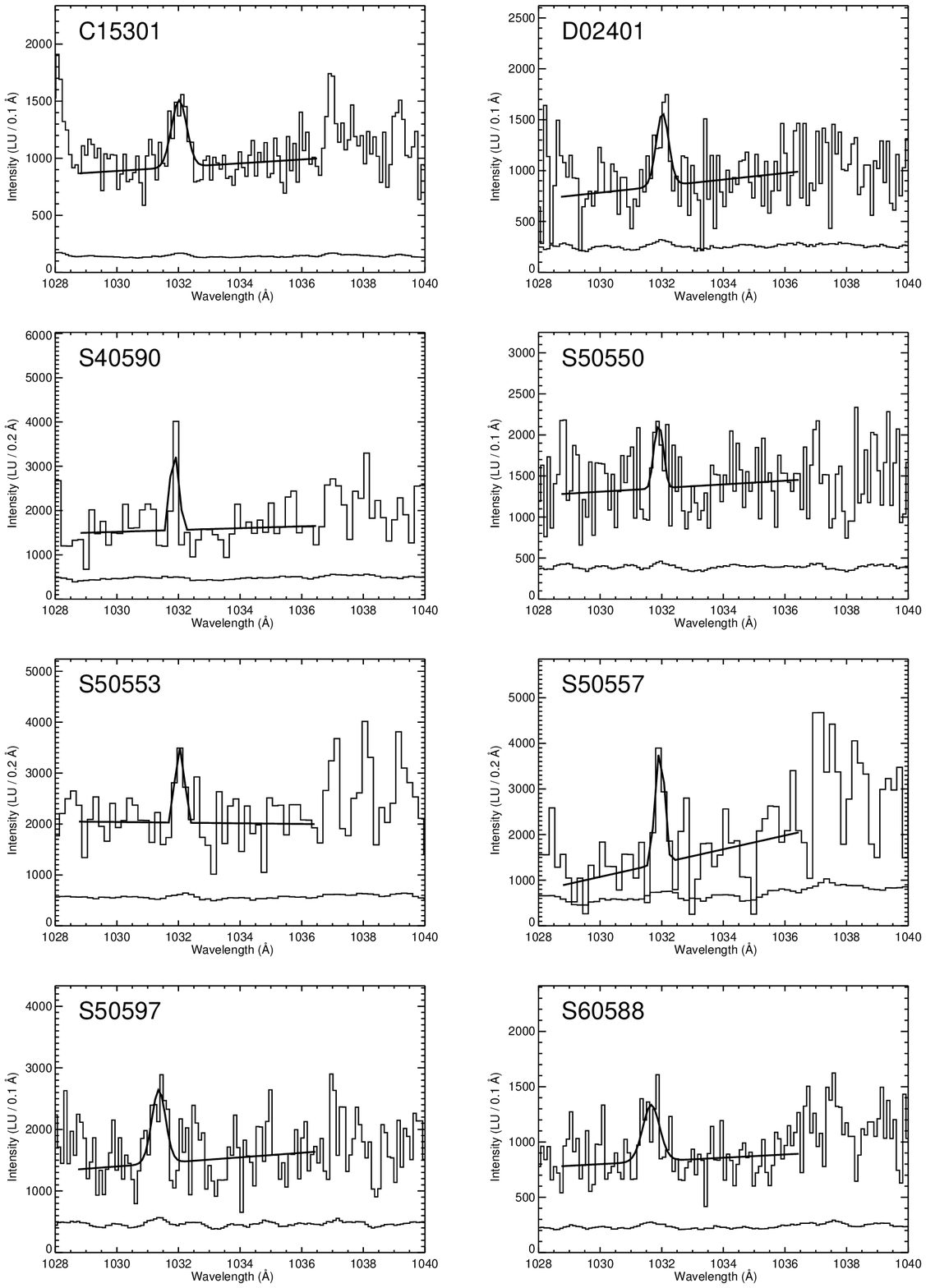}
\caption{
High-significance (3\sig) \osix\ $\lambda 1032$ emission features.
The data are binned by the factors given in Table \ref{tab_detections}, and best-fit model spectra and error bars are overplotted.
Interstellar \ctwo * $\lambda 1037$ and \osix\ $\lambda 1038$ are present in some spectra, as are the geocoronal \oone\ $\lambda \lambda 1028, 1039$ lines.}
\label{fig_3sigma}
\end{figure}

\begin{figure}
\epsscale{.85}
\plotone{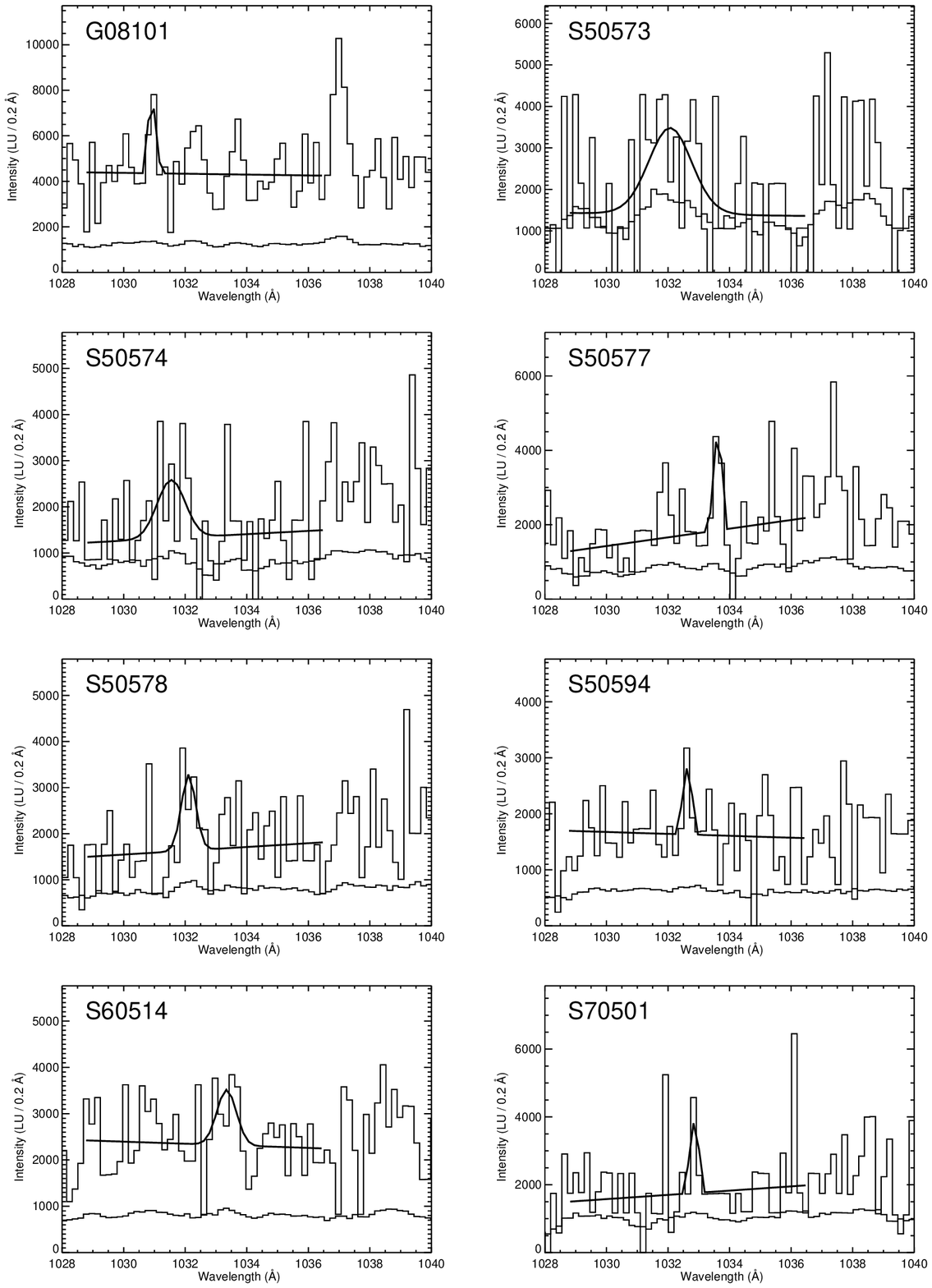}
\caption{Low-significance (2\sig) \osix\ $\lambda 1032$ emission features.
The data are binned by the factors given in Table \ref{tab_detections}, and
best-fit model spectra and error bars are overplotted.
Interstellar \ctwo * $\lambda 1037$ and \osix\ $\lambda 1038$ are present in some spectra,
as are the geocoronal \oone\ $\lambda \lambda 1028, 1039$ lines.}
\label{fig_2sigma}
\end{figure}

Our automated routine flags 32 spectra as having statistically significant \osix\ emission; \ie, the intensity of a putative emission feature is greater than 3 times the uncertainty in the local continuum.  For only 13 of these spectra do model fits return line intensities with $I/\sigma(I) \geq 3$.  We refer to these features as 3\sig\ detections.  Lines with $I/\sigma(I) < 3$ are referred to as 2\sig\ detections.  The best-fit parameters for both sets of \osix\ features are presented in Table \ref{tab_detections}, and plots of the spectra and best-fit models are presented in Figures \ref{fig_3sigma} and \ref{fig_2sigma}.  The asymmetric shapes of the narrowest model emission features are artifacts of our sparse sampling of the model lines.  

Measured intensities range from 2.8 to 10.5 kLU, with a median of 4.0 kLU.  Five of our 3\sig\ detections have local standard of rest velocities $|v_{\rm LSR}| > 60$ \kms, while the rest have $|v_{\rm LSR}| < 35$ \kms.  Four sight lines have FWHM values $< 25$ \kms, indicating that their \osix\ emission does not fill the LWRS aperture.  For these sight lines, the surface brightness of the emitting region is underestimated and the velocity uncertain, as discussed in \citetalias{Dixon:06}.  
For the non-detection sight lines, we compute 3\sig\ upper limits to the intensity of an \osix\ emission feature using the mean background intensity between 1030 and 1035 \AA\ and assuming a line width of 28 pixels (106 \kms).   Upper limits for 15 sight lines with $I(\lambda 1032) \leq 2$ kLU are given in Table \ref{tab_limits}.

Four of our sight lines are previously published, but we include them here because additional data are now available.  The original 26 ks observation of sight line B12901 is analyzed by \citet{Shelton:03} and in \citetalias{Dixon:06}.  Recently, this line of sight was re-observed as target U10635, increasing the total exposure time to 40 ks.  For this line of sight, an upper limit of 1.5 kLU is reported in \citetalias{Dixon:06} (see their Table 6); our new upper limit is 1.2 kLU.  Similarly, sight line C15301 was re-observed as target U10631, increasing the total exposure time from 93 to 102 ks.  Our results (Table \ref{tab_detections}) are consistent with those of \citet{Shelton:07}, who analyze the original C15301 data.  A 5.2 ks observation of sight line S40577 was analyzed in \citetalias{Dixon:06}, resulting in an upper limit of 3.6 kLU (see their Table 4).  The total exposure time is now 24 ks, and our upper limit has fallen to 1.7 kLU.  

Finally, a second observation of sight line S40590 has increased its exposure time from 10.8 to 15.2 ks.  The measured flux is unchanged from that reported in \citetalias{Dixon:06}, but the best-fit radial velocity has risen from $-49$ to $-18$ \kms.  Part of this change may be attributed to the fact that, because the feature is so narrow, we binned the spectrum by 8 pixels in \citetalias{Dixon:06}, while we bin it by 14 pixels for this paper.  Binning the current spectrum by 8 pixels yields a best-fit velocity of $-29 \pm 17$ \kms.  The larger problem is that the emission feature, with FWHM = 9 \kms, is unresolved, suggesting that the emitting region is smaller than the LWRS aperture.  The uncertainty of its position within the aperture adds a systematic velocity uncertainty of about 50 \kms.

\section{Discussion} \label{sec_discussion}

\subsection{Probing the Southern Galactic Sky} \label{sec_map}

Examination of Tables \ref{tab_detections} and \ref{tab_limits} reveals that most of the \osix\ detection sight lines in our sample lie in the region of the southern Galactic sky between $l$ = 180\arcdeg\ and 360\arcdeg, while most of our upper limits lie in the region of the northern Galactic sky between $l$ = 0\arcdeg\ and 180\arcdeg.  The prominence of these two quadrants in the \fuse\/ target list is discussed in \citetalias{Otte:06} and reflects observational constraints that reduced target availability in the orbit plane.  The dearth of \osix\ detections among our northern sight lines may reflect their generally lower galactic latitudes, which are more likely to suffer significant extinction.  Happily, the southern quadrant between $l$ = 180\arcdeg\ and 360\arcdeg\ is now reasonably well sampled in \osix\ emission, and we focus our attention there.

Figure \ref{fig_map} presents a map of the southern Galactic sky centered on the Magellanic Clouds.  The background image is an \hone\ map based on 21 cm emission measurements.  Velocities are color-coded, with purple/blue representing negative velocities and orange/red representing positive velocities.  Overplotted are large circles representing sight lines with significant high-velocity \osix\ absorption from the surveys of \citet{Sembach:03} and Wakker (in preparation).  Their colors indicate the velocity of the absorbing gas using the same color scale as the \hone\ emission.  Small open circles indicate null detections.  The large squares represent 3\sig\ \osix\ detections from this survey and that of \citetalias{Dixon:06}, also color-coded by velocity.  Small open squares represent interesting upper limits ($I(\lambda 1032) \leq 2$ kLU) from both this survey and \citetalias{Dixon:06}.  For convenience, the \osix\ detections and limits taken from \citetalias{Dixon:06} are reproduced in Tables \ref{tab_old_detections} and \ref{tab_old_limits}.

Immediately apparent in \fig{fig_map} are the many sight lines exhibiting low-velocity \osix\ emission (green squares).  Their wide distribution, extending to low galactic latitudes, suggests that they trace a larger structure (\S \ref{sec_compare}).  Among the low-velocity sight lines are two that also exhibit \osix\ absorption (Fairall~9 and NGC~625), from which we can derive the physical parameters of the \osix -bearing gas (\S \ref{sec_parameters}).  Finally, the three sight lines showing high-velocity \osix\ emission (S70504, S90513, and I20505) allow us to probe the Magellanic System (\S \ref{sec_magellanic}).

\subsection{Comparing Low and High Latitudes} \label{sec_compare}

In \citetalias{Dixon:06}, we found evidence for two distinct flavors of \osix -emitting gas: relatively narrow (0.1 pc), high-density (0.1 cm$^{-3}$) conductive interfaces in the local ISM and more extended (100 pc), low-density (0.01 cm$^{-3}$) regions in high-velocity clouds.  To look for evidence of this dichotomy within our current data set, we plot the measured \osix\ emission line intensity, line width, and velocity for the low-velocity 3\sig\ emission features shown in \fig{fig_map} as a function of galactic latitude.  The \osix\ intensities are plotted in \fig{fig_intensity}.  Overplotted is the color excess \ebv\ along each line of sight \citep*{Schlegel:98}.  The figure includes a single sight line, S50508, with an \osix\ intensity of 9.1 kLU and a color excess \ebv\ = 1.26 (off scale).  As discussed in \citetalias{Dixon:06}, this sight line probes the outskirts of the Vela supernova remnant (250 pc distant).  As it is not representative of the general ISM, we will not discuss it further.  

The other sight lines fall into two groups: those at latitudes $|b| > 30\arcdeg$, which suffer little reddening and whose intensities appear to fall with distance from the plane, and those with $|b| < 30\arcdeg$, which suffer significant reddening.  A color excess of \ebv\ = 0.2 corresponds to an attenuation factor of 14 \citep{Fitzpatrick:99}, so it is unlikely that much of the absorbing material lies between us and the emitting gas.  Either these low-latitude sight lines probe relatively nearby gas, or the extinction is patchy -- or both.

The Gaussian FWHM values are plotted in \fig{fig_fwhm}.  This parameter represents the combined effects of the instrumental line-spread function ($\sim$ 25 \kms) and the intrinsic velocity structure of the \osix -emitting gas.  Though the uncertainties are large, the dichotomy between high- and low-latitude sight lines persists, with generally broad emission features at high latitudes and generally narrow features at low latitudes.  It is striking that all four unresolved emission features (those with FWHM $<$ 25 \kms, indicating that they do not fill the 30\arcsec\ $\times$ 30\arcsec\ LWRS aperture) lie at latitudes $|b| < 30\arcdeg$.  If their emitting regions lie within 1 kpc of the earth, then they have a maximum projected spatial extent of 0.15 pc, similar to the path lengths derived for conductive interfaces in the local ISM \citepalias{Dixon:06}.  At temperatures near $3 \times 10^5$ K, thermal broadening would produce line widths of 40 \kms\ (after convolution with the \fuse\/ line-spread function), so the broad emission features, more common at high latitudes, sample either turbulent gas or multiple emitting regions spanning a range of velocities.

The LSR velocities are plotted in \fig{fig_velocity}.  Grey bars represent the range of velocities predicted for each sight line by a simple model of Galactic rotation.  Assuming a differentially-rotating halo with a constant velocity of $v = 220$ \kms\ and a distance from the sun to the Galactic center of 8.5 kpc, we compute the expected radial velocity as a function of distance for the first 5 kpc along each line of sight.  The model ignores broadening due to turbulence or any other motion.  While most of the low-latitude sight lines exhibit \osix\ velocities that are generally consistent with our model, most of the high-latitude sight lines do not.  Instead, most are moving away from the Galactic plane at 10--30 \kms.

These trends are consistent with the picture derived in \citetalias{Dixon:06}: that low-latitude sight lines tend to sample relatively small, dense regions in the local ISM, while high-latitude sight lines sample larger, lower-density, more turbulent regions farther from the disk.  In the following two sections, we quantify the physical properties of these high-latitude emitting regions.

\subsection{\osix\ Column Density} \label{sec_absorption}

Figure \ref{fig_map} reveals \osix\ absorption and emission along two pairs of closely-spaced sight lines: toward the galaxies Fairall~9 and NGC~625.  The Fairall~9 sight lines are more closely spaced.  In an effort to observe \osix\ emission corresponding to the high-velocity \osix\ absorption seen in the spectrum of Fairall~9 \citep{Wakker:03}, sight line D02401 was offset from the galaxy by about 1\arcmin.  NGC~625 was observed in the usual way, with the galaxy in the MDRS aperture and the LWRS aperture observing the background sky some $3\farcm5$ distant.

To measure the \osix\ column densities at velocities corresponding to their observed \osix\ emission, we obtained the fully-reduced \fuse\/ spectra of both galaxies from MAST.  When processing spectra from faint targets, the \fuse\/ data-reduction pipeline runs CalFUSE (version 3.2.1) on each individual exposure, combines the data from all exposures into a single photon-event list, and extracts one spectrum for the whole observation.  This technique maximizes the fidelity of the two-dimensional scattered-light model that is subtracted from the data.  The resulting spectra are plotted in the top panels of Figs.\ \ref{fig_D02401} and \ref{fig_D04001}.  Both sight lines lie in the direction of the Magellanic Stream, and both spectra show \osix\ absorption extending to high velocities.  Following \citet{Wakker:03}, we choose 100 \kms\ as the dividing line between Galactic and Magellanic absorption.

We approximate an LSR velocity scale by shifting the spectra so that the interstellar \ion{Ar}{1} $\lambda 1048.2$  line falls at its laboratory wavelength, then fit a low-order polynomial to the continuum on either side of the \osix\ absorption.  To account for absorption by molecular hydrogen, we scale the model continuum by a synthetic P(3) $\lambda 1031.2$ absorption spectrum using the column densities, velocities, and line widths of \citet{Richter:01}, \citet{Wakker:06}, and B. P. Wakker (2008, private communication).  \htwo\ absorption is seen at velocities of $-1$ and 196 \kms\ toward Fairall~9 and 0 and 396 \kms\ toward NGC~625, as shown in the second panel of Figs.\ \ref{fig_D02401} and \ref{fig_D04001}.

We derive \osix\ column densities using the apparent optical depth method \citep{Savage:Sembach:91}, which is applicable when the spectral line to be measured is unsaturated and nearly resolved.  The third panel of Figs.\ \ref{fig_D02401} and \ref{fig_D04001} shows the resulting column density $N_a$(\osix) as a function of velocity.  (Note that $N_a$ is plotted in units of cm$^{-2}$ per velocity bin.)  The resulting column densities for the low-velocity gas ($-110 < v < 100$ \kms) are $\log N = 14.38$ [cm$^{-2}$] toward Fairall~9 and $\log N = 14.50$ toward NGC~625.  Our result for Fairall~9 is identical to that of \citet{Savage:03}, though they do not correct for \htwo\ absorption in this spectrum.  

The fourth panel of Figs.\ \ref{fig_D02401} and \ref{fig_D04001} reproduces the \osix\ emission spectrum for each sight line.  In both cases, the peak of the \osix\ emission lines up nicely with the low-velocity component of the apparent column density $N_a$(\osix).  We can quantify this alignment by computing the average LSR velocity of the Galactic \osix\ absorption ($-110 < v < 100$) using Equation 3 of \citet{Savage:03}.  For Fairall~9, $\bar{v} = 35$ \kms, consistent with the velocity ($16 \pm 21$ \kms) of the emitting gas.  (\citealt{Savage:03} derive $\bar{v} = 24$ \kms\ over the same velocity range.)  For NGC~625, $\bar{v} = 28$ \kms, again consistent with the velocity ($19 \pm 18$ \kms) of the \osix\ emission.

\subsection{Physical Parameters of the High-Latitude Gas} \label{sec_parameters}

Measurements of \osix\ absorption and emission can be combined to provide valuable diagnostics of the \osix -bearing gas.  Because the \osix\ column density is proportional to the density of the gas, while the line intensity is proportional to the density squared, the ratio of intensity to column density is proportional to the electron density in the emitting plasma.  

The observed \ion{O}{6} $\lambda$1032 intensity toward Fairall~9 is $3.5 \pm 1.1$ kLU.  We assume that the intrinsic intensity of the \ion{O}{6} $\lambda$1038 emission is half that of the 1032 \AA\ line, as would be expected if the emitting gas were optically thin.  The column density of the low-velocity ($-110 < v < 100$) \osix\ is $\log N = 14.38$.  For an assumed temperature of $2.9 \times 10^5$ K, Equation 5 from \citet{Shull:Slavin:94} yields an electron density $n_{\rm e}=9.5 \times 10^{-3}$ cm$^{-3}$ and a thermal pressure $P/k = 1.92 \, n_{\rm e} \, T = 5300$ K cm$^{-3}$.  The color excess toward Fairall~9 is \ebv\ = 0.025 \citep{Schlegel:98}, which attenuates emission at 1032 \AA\ by $\sim$ 30\% \citep{Fitzpatrick:99}.  If all of that material lies between us and the emitting gas, $n_{\rm e}$ rises to $1.3 \times 10^{-2}$ cm$^{-3}$ and $P/k$ to 7400 K cm$^{-3}$.  

To calculate the \ion{O}{6} density and the path length through the emitting gas, we need the oxygen abundance and the fraction of oxygen in O$^{+5}$.  From their sample of low-density ($\langle n_{\rm H} \rangle < 1.0$ cm$^{-3}$) Galactic sight lines longer than 800 pc, \citet{Cartledge:04} derive a mean interstellar O/H ratio of $(4.37 \pm 0.09) \times 10^{-4}$.  For plasmas in collisional ionization equilibrium, the O$^{+5}$ fraction peaks at 21.5\% when the gas temperature is $2.9 \times 10^5$ K \citep{Gnat:Sternberg:07}.  With these values, and assuming that the gas is completely ionized ($n_{\rm e} = 1.2 n_{\rm H}$), we derive an \ion{O}{6} density of $7.4 \times 10^{-7}$ cm$^{-3}$ and a path length through the gas of 100 pc for zero reddening.  If \ebv\ = 0.025, then the \ion{O}{6} density and path length become $1.0 \times 10^{-6}$ cm$^{-3}$ and 75 pc, respectively.

The observed \ion{O}{6} $\lambda$1032 intensity toward NGC~625 is $2.6 \pm 0.8$ kLU and the column density of its low-velocity gas is $\log N$ = 14.50.  From these values, we derive an electron density for the \osix -bearing gas of $n_{\rm e}=5.3 \times 10^{-3}$ cm$^{-3}$, a thermal pressure $P/k = 3000$ K cm$^{-3}$, an \ion{O}{6} density of $4.2 \times 10^{-7}$ cm$^{-3}$, and a path length through the emitting gas of 250 pc.  The color excess toward NGC~625 is \ebv\ = 0.016, which attenuates emission at 1032 \AA\ by $\sim$ 20\%.  Correcting for this attenuation raises the density and pressure and reduces the path length by the same factor.

\citet{Shelton:07} use shadowing arguments to conclude that essentially all of the \osix\ emission seen along sight line C15301 (at $l = 278.7, b = -47.1$ in \fig{fig_map}) comes from the Galactic halo.  Lacking a measurement of the \osix\ column density along this line of sight, they adopt an average of the published measurements for four nearby sight lines, $\log N$ = 14.37, from which they derive an electron density $n_{\rm e}$ = 1.1--1.6 $\times 10^{-2}$ cm$^{-3}$,  a thermal pressure $P/k = $7--10$ \times 10^3$ K cm$^{-3}$, and a path length of 50--70 pc for the emitting gas.  These parameters are consistent with those derived from the Fairall~9 and NGC~625 data, suggesting a thick disk/halo origin for the \osix -emitting gas along all three lines of sight.

We can derive a lower limit to the distance of the \osix -emitting gas from the angular distance between a pair of \osix\ detection and upper-limit sight lines that are near one another on the sky.  The two closest pairs are I20505/B12901 and C15301/Z90719, but both probe variations in the opacity of an intervening dust cloud (see the discussion in Appendix A2 of \citetalias{Dixon:06}), rather than variations in the intrinsic \osix\ emission.  The third closest pair, C07601/S90522, is separated by $\sim$ 3\arcdeg.  The non-detection of \osix\ along sight line S90522 is not due to interstellar opacity; the extinction maps of \citet{Schlegel:98} suggest that it is less affected by dust than is C07601.  If the \osix -emitting region sampled by sight line C07601 has a radius of $\sim$ 100 pc (assuming size $\sim$ path length), then it lies at least 2 kpc from the Sun and at least 800 pc from the Galactic plane.  

\subsection{Comparison with Models} \label{sec_models}

A successful model of the hot component of the interstellar medium must reproduce our observational results:  Diffuse \osix\ emission (\ie, emission unrelated to discrete supernova remnants or superbubbles) is present over the entire sky with intensities between about 2 and 6 kLU.  The emission is patchy: of the sight lines providing statistically-significant results, roughly 40\% yield upper limits \iosix\ $< 2$ kLU.  Near the plane, the emitting regions are less than about 0.15 pc in spatial extent, with electron densities $n_{\rm e} = 0.2-0.3$ cm$^{-3}$  (\citetalias{Dixon:06}).  At high latitudes, the emitting regions are large, with path lengths on the order of 100 pc, and at least 800 kpc from the plane.  Electron densities are on the order of $10^{-2}$ cm$^{-3}$ 

In a series of papers, \citet{deAvillez:04, deAvillez:05, deAvillezOVI} present the results of high-resolution, three-dimensional hydrodynamic and magnetohydrodynamic simulations of the ISM.  The authors model a section of the Galaxy centered on the solar circle having an area of 1 kpc$^2$ and extending $\pm$ 10 kpc from the Galactic plane.  They employ an adaptive mesh refinement algorithm whose resolution peaks at 1.25 pc within the disk ($|z| \leq 500$ pc).  By computing \nosix\ along hundreds of sight lines (all with $|z| \leq 250$ pc and $d \leq 1$ kpc) through their data cube, \citet{deAvillezOVI} are able to reproduce both the observed distribution of \nosix\ with distance and its dispersion.

We derive electron densities $n_{\rm e} \sim 10^{-2}$ cm$^{-3}$ for the \osix-emitting gas along the lines of sight toward Fairall~9 and NGC~625.  \citet[see their Fig.\ 2]{deAvillez:05} present a two-dimensional vertical slice (perpendicular to the midplane) through their three-dimensional data cube showing the density distribution of the ISM.  Gas parcels with densities on the order of $10^{-2}$ cm$^{-3}$ are present throughout the frothy thick disk of ionized and neutral gas lofted by supernova explosions in the plane.  The boundaries of these parcels would be natural locations for the formation of \osix -bearing interfaces.  Model predictions of \nosix\ and \iosix\ as functions of Galactic latitude have not yet been published.

\subsection{\osix\ Emission from the Magellanic System} \label{sec_magellanic}

The three sight lines in \fig{fig_map} with high-velocity \osix\ emission are S70504 (+84 \kms) and S90513 (+115 \kms) from the current sample and I20505 (+206 \kms) from \citetalias{Dixon:06}.  All three lie in a region of the Magellanic System designated as the ``interface region'' by \citet[see their Fig.\ 4]{Bruns:05}.  The interface region connects the Magellanic Clouds and the Magellanic Bridge with the Magellanic Stream, which lies to the south of $-60\arcdeg$ latitude.  The \hone\ in this region exhibits a complex filamentary structure and a distinct velocity signature.  Sight lines S90513 and I20505 lie near the
projected western and eastern edges of the region, respectively, and the observed
LSR velocities of their \osix\ emission closely match those of the \hone.
Although the \osix\ velocity along sight line S70504 is about half that of the \hone\ in the same direction, it is within the overall range of \hone\ velocities detected in the interface region, 60--360 \kms.

The interface region of the Magellanic System lies about 40 kpc from the Galactic plane \citep{Jin:08} and is thus embedded in the hot ($10^6$ K) Galactic corona.  The perimeters of these high-velocity \hone\ clouds, where cool and hot gas meet and mingle, are natural sites for the formation of intermediate-temperature \osix -bearing gas.  Indeed, \cite{Sembach:03} report \osix\ absorption along several extragalactic sight lines (including Fairall~9) that probe the Magellanic Stream at velocities consistent with that of the \hone\ gas.  Intriguingly, while both Fairall~9 and NGC~625 show \osix\ absorption at Magellanic velocities, neither shows high-velocity \osix\ emission.  The interface region, which does show \osix\ emission, is unique in this respect.

These are the first detections of high-ionization emission from the Magellanic System outside of the Clouds themselves.  The \osix\ intensities observed along the two sight lines towards the edges of the interface region are well within the normal range observed in the diffuse ISM, but the intensity along sight line S70504 (10.5 kLU) is significantly higher.  In fact, it is the highest diffuse \osix\ intensity observed to date (outside of supernova remnants, superbubbles, and high-ionization planetary nebulae).  Though the \osix\ emission feature in this spectrum is statistically significant, its exposure time is quite short, just over 2 ks.  Deeper and more detailed observations would help to determine whether such high intensities are commonly obtained in the interface region of the Magellanic System.

\section{Conclusions}\label{sec_conclusions}

We present the final installment of the \fuse\/ survey of diffuse \osix\ emission in the interstellar medium.  These observations reveal the large-scale structure of the \osix -bearing gas in the quadrant of the sky centered on the Magellanic Clouds.  Its most prominent feature, a layer of low-velocity \osix\ emission extending more than 70\arcdeg\ from the Galactic plane, is produced by two distinct populations of emitting clouds.  At latitudes $|b| < 30\arcdeg$, sight lines suffer significant extinction, suggesting that they probe relatively nearby emitting regions; their emission features are more likely to be narrow or unresolved, suggesting single interfaces less than about 0.15 pc across; and their velocities are generally consistent with Galactic rotation.  We identify these emitting regions with narrow, high-density interfaces in the local ISM.

At latitudes $|b| > 30\arcdeg$, extinction is low.  Intensities fall with distance from the plane.  Emission features are generally broad, suggesting that they sample turbulent gas or multiple emitting regions spanning a range of velocities.  The \osix -bearing gas seems to be moving away from the Galactic plane.  For two sight lines, Fairall~9 and NGC~625, we combine \osix -emission and -absorption measurements to constrain the physical parameters of the emitting regions.  From the resulting path lengths, we estimate that these regions lie at least 800 pc from the Galactic plane.  The presence of nearby sight lines yielding null detections but strong upper limits suggests that the emission is patchy, consistent with the results of \citet{Howk:02}.

High-velocity \osix\ emission is seen along three sight lines that probe the interface region of the Magellanic System.  The \osix\ velocities are generally consistent with those of the underlying \hone.  High-velocity \osix\ emission is not seen in sight lines that probe the Magellanic Stream, suggesting that the dynamics of the interface region differ from those of the Stream itself.  These are the first detections of emission from high-ionization species in the Magellanic System outside of the Clouds themselves.

\acknowledgments

\begin{figure}
\includegraphics[scale=0.5,angle=270, trim = 80 0  0 0]{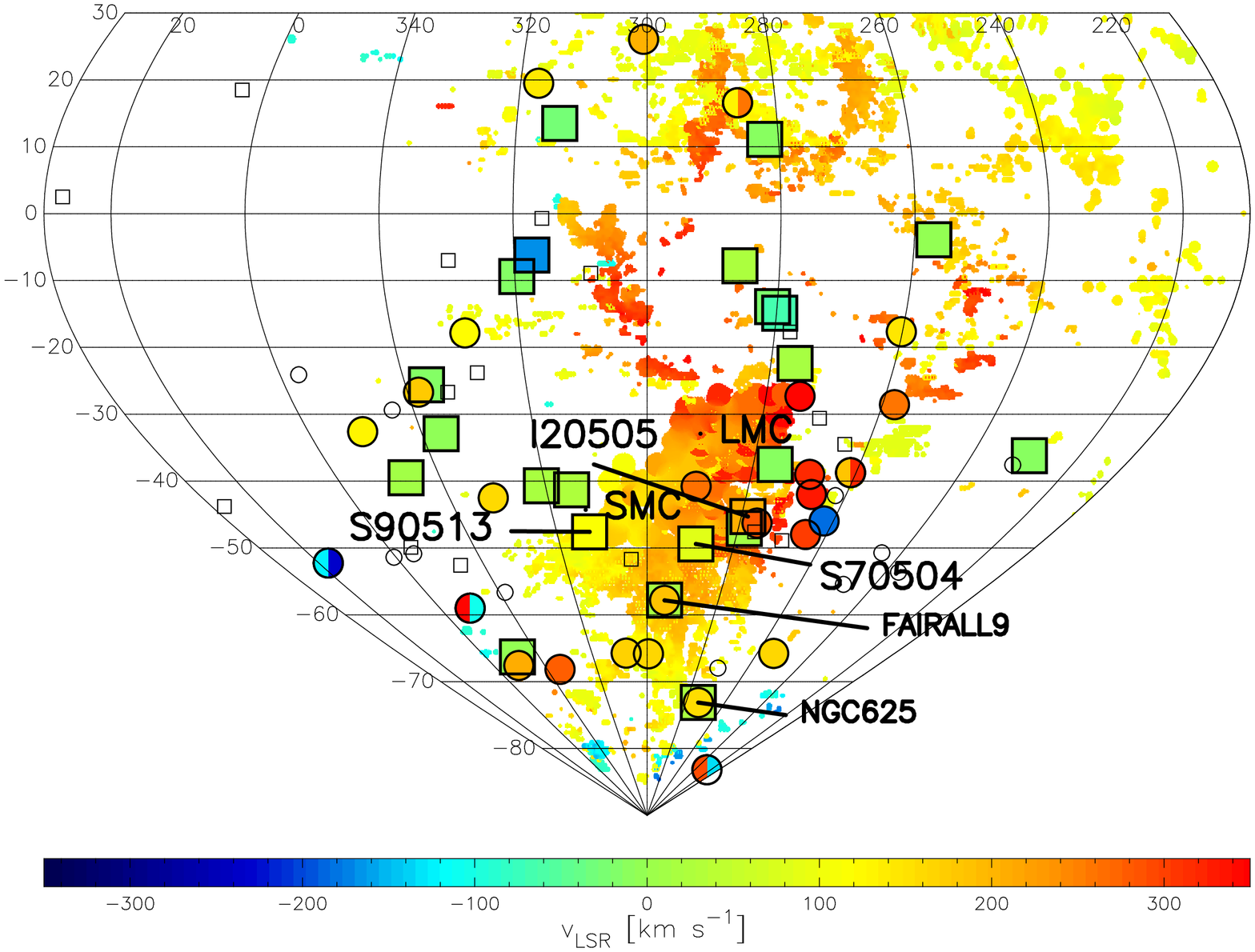}
\caption{Hammer-Aitoff projection of the high-velocity \hone\ sky based on the 21 cm emission measurements of \citet{Morras:00} and \citet{Hulsbosch:88}.  The \hone\ data have a spatial resolution of approximately 36\arcmin\ and are representative of gas with \nho\ $ > 2 \times 10^{18}$ cm$^{-2}$.  The map is centered on the Magellanic Clouds; coordinates are galactic longitude and latitude.
Positions of the high-velocity \osix\ absorption features reported in \citet{Sembach:03} and Wakker (in preparation) are indicated by the large circles, whose colors indicate velocity on the same color scale used for the \hone\ emission. If more than one high-velocity \osix\ absorption feature is present along a line of sight, the circle is split and the velocities are color-coded in each section of the circle. Small open circles indicate null detections.
Positions of the \osix\ emission features reported in this paper and in \citetalias{Dixon:06} are indicated by large squares, also color-coded by velocity.  Small squares represent upper limits $\leq$ 2 kLU.  Labeled sight lines are discussed in the text.}
\label{fig_map}
\end{figure}

\begin{figure}
\epsscale{.75}
\plotone{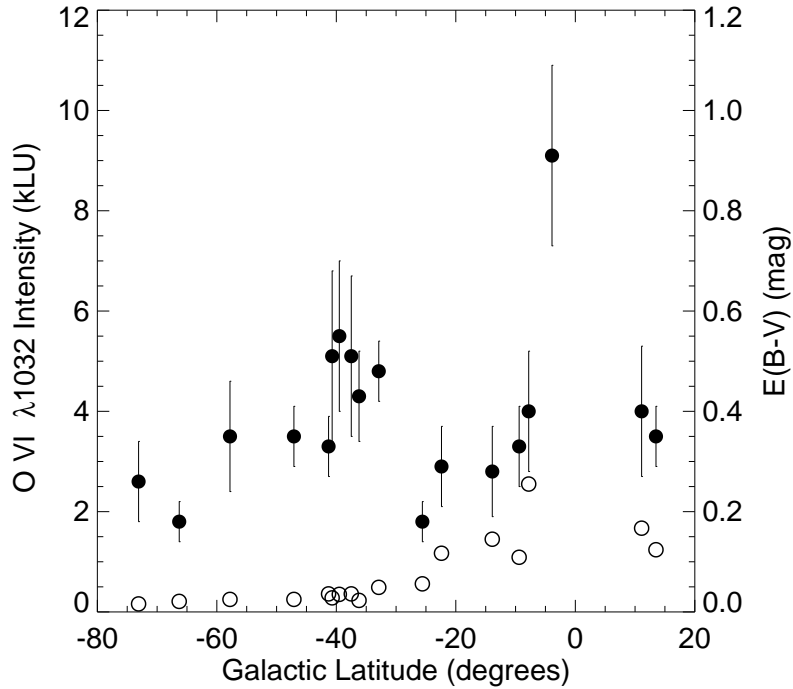}
\caption{Variation of observed \osix\ $\lambda 1032$ intensity with galactic latitude.  Solid circles represent the low-velocity ($|v_{\rm LSR}| < 60$ \kms) 3\sig\ detections from \fig{fig_map}.  Open circles represent the color excess \ebv\ along each line of sight. Note that the color excess is scaled by a factor of ten relative to intensity.}
\label{fig_intensity}
\end{figure}

\begin{figure}
\epsscale{.75}
\plotone{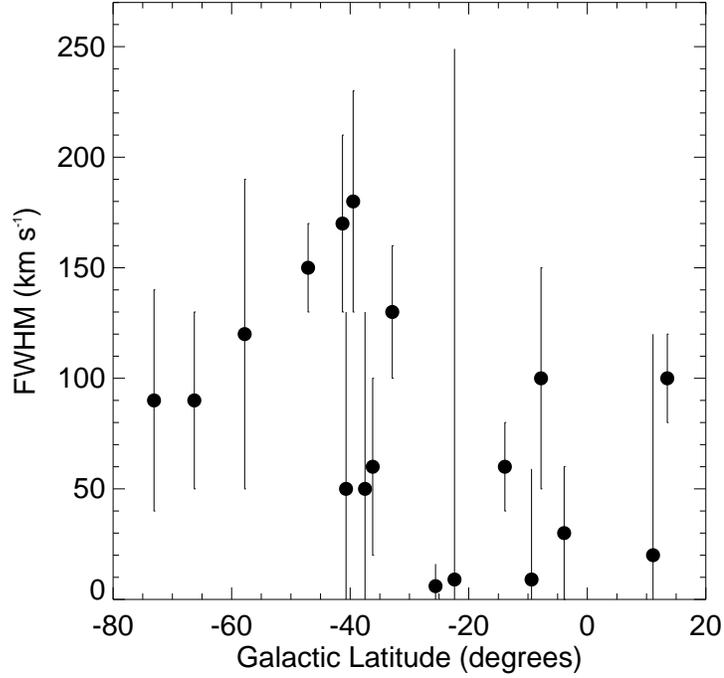}
\caption{Best-fit Gaussian FWHM value as a function of galactic latitude.  Solid circles represent the low-velocity ($|v_{\rm LSR}| < 60$ \kms) 3\sig\ detections from \fig{fig_map}.}
\label{fig_fwhm}
\end{figure}

\begin{figure}
\epsscale{.75}
\plotone{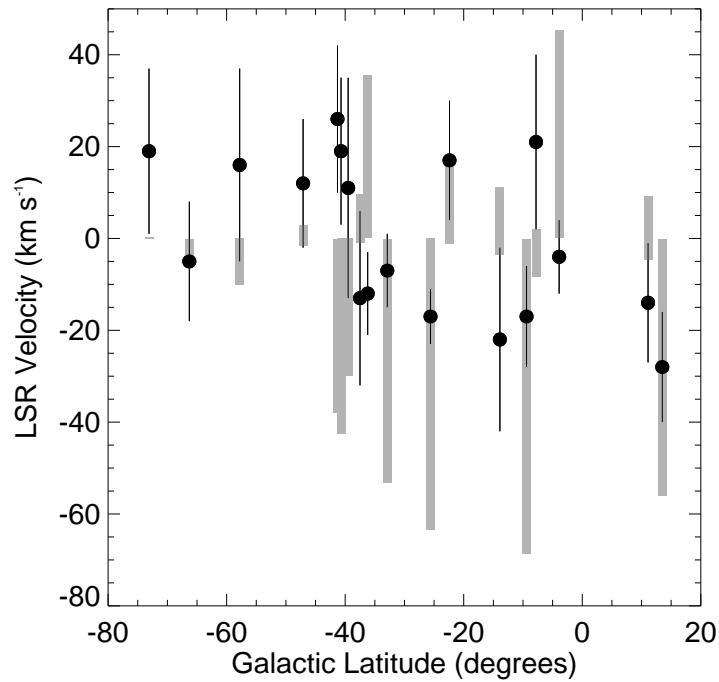}
\caption{
Variation of LSR velocity with galactic latitude.  Solid circles represent the low-velocity ($|v_{\rm LSR}| < 60$ \kms) 3\sig\ detections from \fig{fig_map}.  Grey bars represent the range of LSR velocities predicted for the first 5 kpc along each sight line by a simple (cylindrical, constant-velocity) model of Galactic rotation.}
\label{fig_velocity}
\end{figure}

\begin{figure}
\epsscale{.75}
\plotone{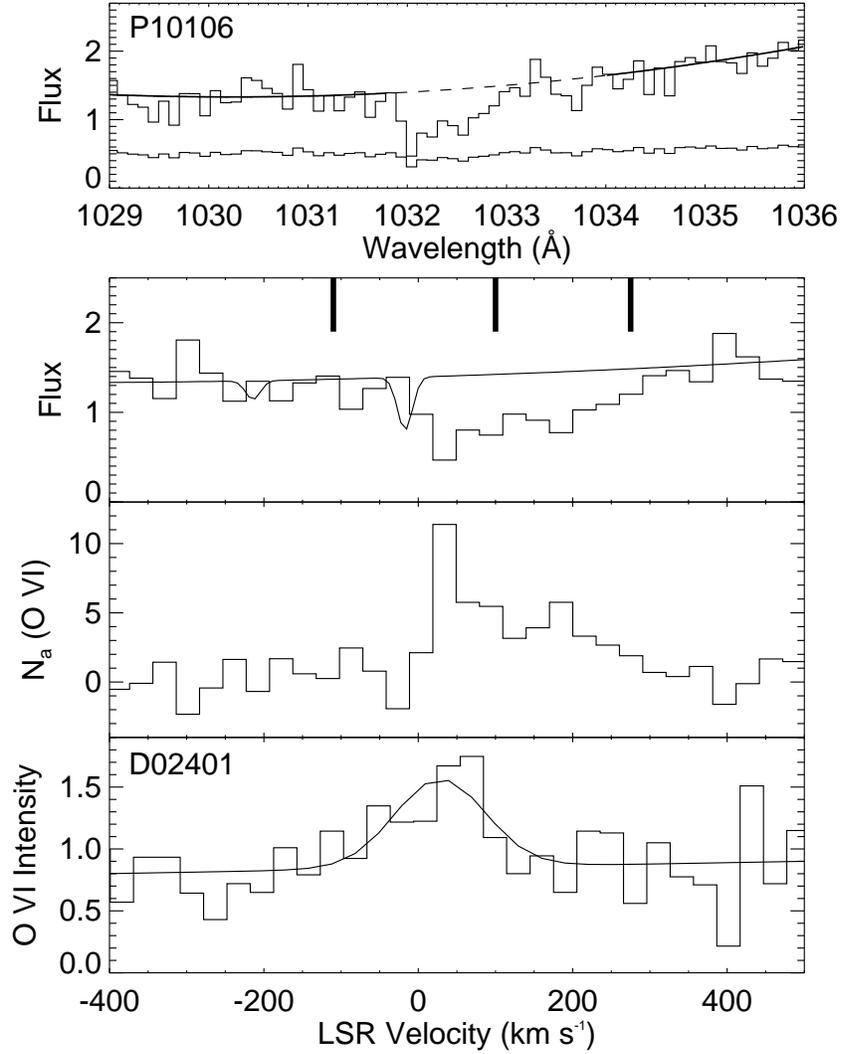}
\caption{{\it First panel:}\/ \fuse\/ spectrum of Fairall 9 (P10106) showing \osix\ $\lambda 1032$ absorption from both the Galaxy and the Magellanic Stream.  Overplotted are a continuum fit (interpolated over the dashed region) and the error array.  Flux units are \flux.  {\it Second panel:}\/  Spectrum of Fairall 9 plotted as a function of LSR velocity.  The spectrum is binned by 8 pixels, or 30 \kms.  The continuum is scaled to account for absorption from the R(3) $\lambda 1031.19$ line of molecular hydrogen (at both low and high velocities).  Thick vertical lines mark the limits assumed for the Galactic ($-110 < v < 100$ \kms) and Magellanic Stream ($100 < v < 275$ \kms) absorption.  {\it Third panel:}\/ Apparent \osix\ column density in units of $10^{13}$ cm$^{-2}$ per velocity bin, computed as described in the text.  {\it Fourth panel:}\/ \osix\ emission in units of kLU per 30 \kms\ velocity bin.}
\label{fig_D02401}
\end{figure}

\begin{figure}
\epsscale{.85}
\plotone{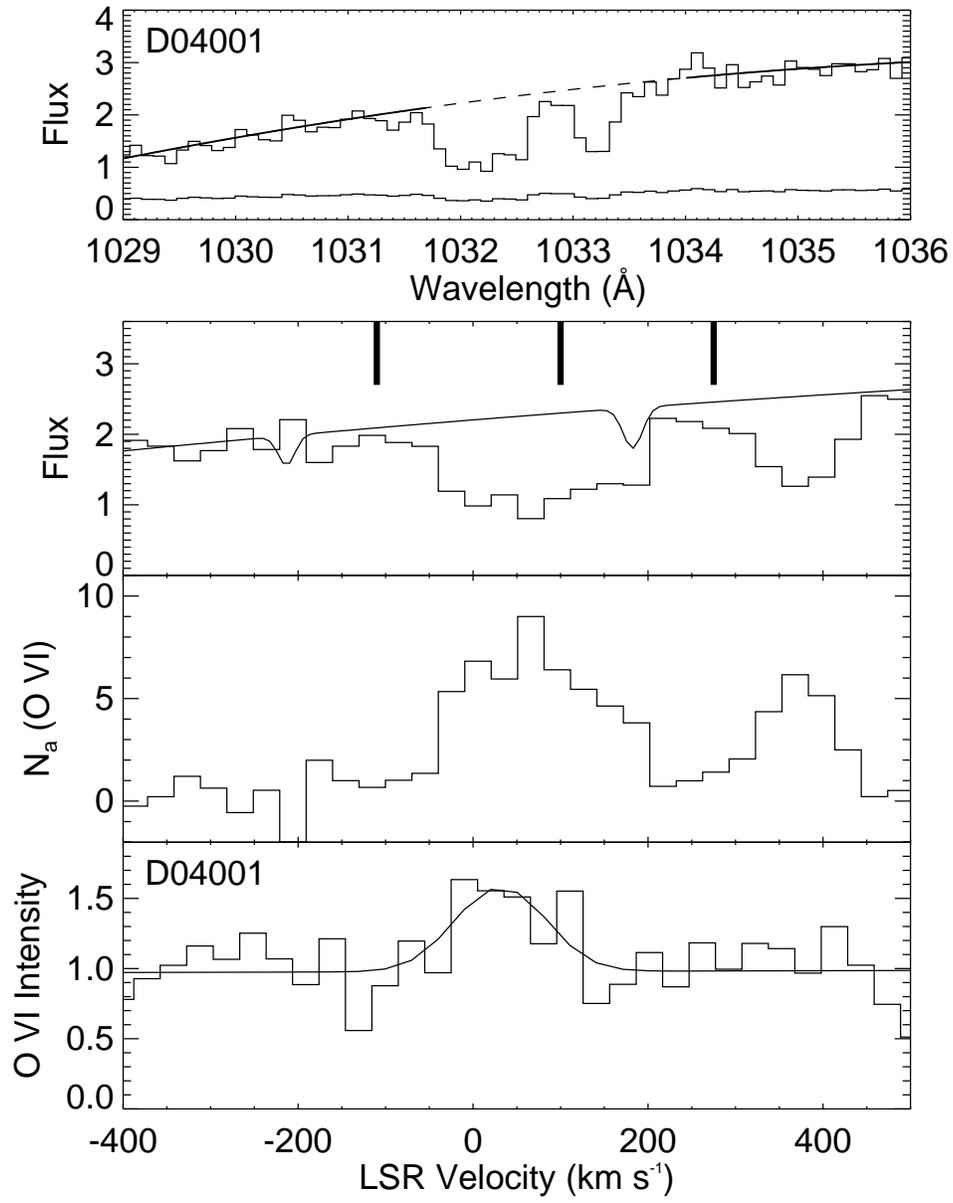}
\caption{Same as \fig{fig_D02401}, but for NGC~625 (D02401).  The \osix\ emission spectrum is from \citetalias{Dixon:06}.}
\label{fig_D04001}
\end{figure}

The authors would like to thank B. P. Wakker for constructing Figure \ref{fig_map}, for providing \osix\ absorption-line data in advance of publication, and for helpful comments on the text.  We acknowledge with gratitude the extraordinary efforts of the \fuse\/ operations team to make this mission successful.  This research has made use of the NASA Astrophysics Data System (ADS) and the Multimission Archive at the Space Telescope Science Institute (MAST).  STScI is operated by the Association of Universities for Research in Astronomy, Inc., under NASA contract NAS5-26555. Support for MAST for non-HST data is provided by the NASA Office of Space Science via grant NAG5-7584 and by other grants and contracts.  
The curve-fitting program \specfit\ runs in the IRAF environment.
The Image Reduction and Analysis Facility is distributed by the National Optical Astronomy Observatories, which is supported by the Association of Universities for Research in Astronomy (AURA), Inc., under cooperative agreement with the National Science Foundation.
This work is supported by NASA grant NAS5-32985 to the Johns Hopkins University.

{\it Facilities:} \facility{FUSE}.


\clearpage

\begin{deluxetable}{llrrrrrcrrc} 
\tablecolumns{11} 
\tablewidth{0pt}
\tabletypesize{\scriptsize}
\tablecaption{\label{tab_detections} \ion{O}{6} $\lambda 1032$ Detections}
\tablehead{ 
\colhead{Sight}  & \colhead{Target} & \colhead{$l$}   & \colhead{$b$}   & \colhead{$t$\tablenotemark{b}} & & \colhead{Intensity\tablenotemark{d}} & \colhead{Wavelength\tablenotemark{e}} &
\colhead{FWHM\tablenotemark{f}}   & \colhead{$v_{\rm LSR}$} & \colhead{$E(B-V)$\tablenotemark{g}} \\
\colhead{Line}    & \colhead{Name\tablenotemark{a}} & \colhead{(deg)} & \colhead{(deg)} & \colhead{(s)} & \colhead{Bin\tablenotemark{c}} & \colhead{(kLU)}   & \colhead{(\AA)} &
\colhead{(km s$^{-1}$)} & \colhead{(km s$^{-1}$)}   & \colhead{(mag)} }
\startdata 
\cutinhead{$3 \sigma$ Detections} 
C15301\tablenotemark{h} & SKY-032010-622629 & 278.7 & $-47.1$ & 102371 &  8 & $3.5 \pm 0.6$ & $1032.02 \pm 0.05$ & $150 \pm 20$ & $  12 \pm  14$ & 0.025\\
D02401 & FAIRALL9-OFF      & 295.1 & $-57.8$ &  28325 &  8 & $3.5 \pm 1.1$ & $1032.02 \pm 0.07$ & $120 \pm 70$ & $  17 \pm  21$ & 0.025\\
S40590 & HE2-138-BKGD      & 319.7 & $ -9.4$ &  15165 & 14 & $3.3 \pm 0.8$ & $1031.88 \pm 0.04$ & $  9 \pm 50$ & $ -18 \pm  11$ & 0.109\\
S50550 & BKGD              & 280.7 & $-13.9$ &  18453 &  8 & $2.8 \pm 0.9$ & $1031.90 \pm 0.07$ & $ 60 \pm 20$ & $ -22 \pm  20$ & 0.145\\
S50553 & BKGD              & 276.1 & $-22.4$ &  13595 & 14 & $2.9 \pm 0.8$ & $1032.04 \pm 0.04$ & $  9 \pm240$ & $  18 \pm  13$ & 0.117\\
S50557 & BKGD              & 275.9 & $-37.5$ &   7815 & 14 & $5.1 \pm 1.6$ & $1031.94 \pm 0.06$ & $ 50 \pm 80$ & $ -13 \pm  19$ & 0.036\\
S50597 & BKGD              & 317.3 & $ -6.2$ &  15359 &  8 & $6.5 \pm 2.0$ & $1031.36 \pm 0.08$ & $140 \pm 60$ & $-168 \pm  25$ & 0.420\\
S60588 & NBKGD88           &  91.6 &   18.7  &  33917 &  8 & $3.4 \pm 1.0$ & $1031.66 \pm 0.08$ & $180 \pm 60$ & $-61 \pm 23$ & 0.068\\
S70504 & SBKGD4            & 288.8 & $-49.4$ &   2223 & 14 & $10.5 \pm 3.1$ & $1032.26 \pm 0.05$ & $ 20 \pm 90$ & $  84 \pm  14$ & 0.024\\
S70512 & SBKGD12           & 279.5 & $-14.9$ &  11214 & 14 & $6.9  \pm 2.0$ & $1031.75 \pm 0.07$ & $140 \pm 60$ & $ -65 \pm 22$ & 0.151\\
S70514 & SBKGD14           & 286.0 &  $-7.8$ &  19190 & 14 & $4.0  \pm 1.2$ & $1032.04 \pm 0.06$ & $100 \pm 50$ & $  22 \pm 19$ & 0.255\\
S90513 & SBKGD             & 312.7 & $-47.6$ &   8609 & 14 & $5.5 \pm 1.2$ & $1032.35 \pm 0.02$ & $  2 \pm 20$ & $ 116 \pm  \phn6$ & 0.028\\
S90523 & SBKGD23A          & 320.8 & $-40.7$ &   8045 & 14 & $5.1 \pm 1.7$ & $1032.01 \pm 0.06$ & $ 50 \pm 80$ & $  19 \pm  16$ & 0.028\\
\cutinhead{$2 \sigma$ Detections} 
G08101 & HD128620          & 315.7 & $ -0.7$ &   5717 & 14 & $5.7 \pm 2.3$ & $1030.92 \pm 0.06$ & $  6 \pm 80$ & $-295 \pm  18$ & 6.505\\
S50573 & BKGD              & 288.0 & $ -6.6$ &   1918 & 14 & $20.3 \pm 8.9$ & $1032.07 \pm 0.32$ & $480 \pm 150$ & $ 29 \pm 94$ &  0.369\\
S50574 & POLE-174-BKG      & 323.1 & $-45.8$ &   4738 & 14 & $7.9 \pm 4.4$ & $1031.54 \pm 0.25$ & $300 \pm 80$ & $-118 \pm  73$ & 0.025\\
S50577 & POLE-177-BKG      & 328.9 & $-39.6$ &   5490 & 14 & $4.8 \pm 1.8$ & $1033.61 \pm 0.06$ & $  2 \pm600$ & $ 485 \pm  17$ & 0.032\\
S50578 & POLE-178-BKG      & 329.7 & $-37.1$ &   5736 & 14 & $5.3 \pm 2.5$ & $1032.11 \pm 0.13$ & $140 \pm 80$ & $  49 \pm  39$ & 0.038\\
S50594 & BKGD              & 129.6 & $ 15.9$ &   8266 & 14 & $2.4 \pm 1.2$ & $1032.62 \pm 0.08$ & $ 30 \pm 60$ & $ 210 \pm  23$ & 0.285\\
S60514 & NBKGD14           & 146.2 & $ 43.7$ &   7341 & 14 & $5.2 \pm 2.1$ & $1033.35 \pm 0.15$ & $200 \pm 50$ & $ 418 \pm  43$ & 0.050\\
S70501 & SBKGD1            & 311.6 & $-51.2$ &   3487 & 14 & $4.1 \pm 2.2$ & $1032.85 \pm 0.08$ & $ 20 \pm140$ & $ 259 \pm  24$ & 0.019\\
S70509 & SBKGD9            & 274.6 & $-28.7$ &  14400 & 14 & $2.1 \pm 1.3$ & $1032.40 \pm 0.09$ & $  7 \pm200$ & $ 121 \pm  27$ & 0.064\\
S70516 & SBKGD16           & 294.9 & $ -3.3$ &   8219 & 14 & $3.5 \pm 1.4$ & $1031.59 \pm 0.06$ & $ 20 \pm270$ & $-109 \pm  18$ & 1.260\\
S70522 & SBKGD22           & 322.2 & $-9.9$  &  23374 & 14 & $1.4 \pm 0.6$ & $1031.65 \pm 0.03$ & $  3 \pm  30$ & $-82 \pm 10$ & 0.137\\
S70527 & SBKGD27           & 331.2 & $-32.0$ &   8016 & 14 & $7.8 \pm 3.8$ & $1032.42 \pm 0.17$ & $200 \pm 90$ & $ 141 \pm  49$ & 0.087\\
S70528 & SBKGD28           & 330.4 & $-36.7$ &  15182 &  8 & $2.9 \pm 1.2$ & $1034.03 \pm 0.05$ & $  6 \pm 30$ & $ 607 \pm  14$ & 0.039\\
S70540 & SBKGD40           & 281.8 & $-21.6$ &   2820 & 14 & $8.3 \pm 3.9$ & $1033.21 \pm 0.08$ & $ 20 \pm210$ & $ 357 \pm  24$ & 0.176\\
S70549 & SBKGD49           & 318.6 & $-13.5$  &  3081 & 14 & $4.2 \pm 2.6$ & $1031.92 \pm 0.09$ & $ 10 \pm 220$ & $ -5 \pm 26$ & 0.106\\
S70567 & SBKGD67           & 268.8 & $-30.6$ &   2001 & 14 & $5.7 \pm 3.6$ & $1030.92 \pm 0.08$ & $  4 \pm 80$ & $-309 \pm  24$ & 0.051\\
S70569 & SBKGD69           & 270.7 & $-21.0$ &   5650 & 14 & $4.8 \pm 2.3$ & $1034.09 \pm 0.07$ & $ 10 \pm230$ & $ 611 \pm  21$ & 0.131\\
S90510 & SBKGD10A          & 274.3 & $-20.4$ &   3445 & 14 & $6.8 \pm 2.9$ & $1032.30 \pm 0.07$ & $  5 \pm 90$ & $  93 \pm  21$ & 0.138\\
S90524 & SBKGD24A          & 278.5 & $-54.3$ &   8968 & 14 & $3.1 \pm 1.2$ & $1031.96 \pm 0.07$ & $ 20 \pm 80$ & $  -4 \pm  19$ & 0.030

\enddata 
\tablenotetext{a}{Target names are taken from the data file headers.}
\tablenotetext{b}{Exposure time is night only.}
\tablenotetext{c}{Spectral binning in 0.013 \AA\ pixels.}
\tablenotetext{d}{kLU = 1000 photon s$^{-1}$\,cm$^{-2}$\,sr$^{-1}$.}
\tablenotetext{e}{Wavelengths are heliocentric.}
\tablenotetext{f}{Gaussian FWHM values include the smoothing imparted by the instrument optics. Values less than $\sim$ 25 \kms\ indicate that the emission does not fill the LWRS aperture.}
\tablenotetext{g}{Extinction from \citealt{Schlegel:98}.}
\tablenotetext{h}{Combines data sets C15301 and U10631.  See discussion in text.}

\end{deluxetable} 

\clearpage

\begin{deluxetable}{llrrrcc}
\tablecolumns{7} 
\tablewidth{0pt}
\tablecaption{$3\sigma$ Upper Limits on \osix\ $\lambda 1032$ Emission \label{tab_limits}}
\tablehead{
\colhead{Sight} & \colhead{Target} & \colhead{$l$} & \colhead{$b$} & \colhead{$t$\tablenotemark{b}} & \colhead{$3\sigma$ Limit\tablenotemark{c}} & \colhead{$E(B-V)$\tablenotemark{d}} \\
\colhead{Line} & \colhead{Name\tablenotemark{a}} & \colhead{(deg)} & \colhead{(deg)} & \colhead{(s)} & \colhead{(kLU)}  & \colhead{(mag)} }

\startdata
A15101 &         IPM-Upwind &   3.7 &  $ 18.5  $ &  24140 & 1.6  & 	0.853 \\
B12901\tablenotemark{e} &  SKY-033255-632751 & 278.6 &  $ -45.3 $  &  39983 & 1.2  & 	0.160 \\
S40577\tablenotemark{f} &     HDE232522-BKGD & 130.7 &  $  -6.7 $  &  24329 & 1.7  & 	0.278 \\
S50551 &               BKGD & 277.6 &  $ -17.7 $  &  30521 & 1.6  & 	0.172 \\
S50589 &           POLE-BKG & 308.5 &  $  -8.9 $  &  33483 & 1.5  & 	0.268 \\
S60503 &             NBKGD3 &  27.3 &  $  2.5  $ &  27071 & 1.5  & 	2.436 \\
S60504 &             NBKGD4 & 132.0 &  $  3.7  $ &  31556 & 1.4  & 	1.159 \\
S60507 &             NBKGD7 & 144.1 &  $ 11.9  $ &  19715 & 1.8  & 	0.512 \\
S60537 &            NBKGD37 & 140.6 &  $ 15.6  $ &  22511 & 1.7  & 	0.199 \\
S60538 &            NBKGD38 & 143.3 &  $ 19.8  $ &  27384 & 1.6  & 	0.108 \\
S60563 &            NBKGD63 & 140.0 &  $  2.0  $ &  14849 & 2.0  & 	1.984 \\
S60565 &            NBKGD65 & 147.6 &  $  8.2  $ &  23516 & 1.6  & 	0.662 \\
S70502 &             SBKGD2 & 303.8 &  $ -51.7 $  &  18927 & 1.8  & 	0.019 \\
S80502 &            NBKGD02 &  90.7 &  $  21.0 $  &  10074 & 2.0  & 	0.074 \\
S90522 &           SBKGD22A & 333.3 &  $ -26.7 $  &  20974 & 2.0 & 	0.047
\enddata

\tablenotetext{a}{Target names are taken from the data file headers.}
\tablenotetext{b}{Exposure time is night only.}
\tablenotetext{c}{kLU = 1000 photon s$^{-1}$\,cm$^{-2}$\,sr$^{-1}$.}
\tablenotetext{d}{Extinction from \citealt{Schlegel:98}.}
\tablenotetext{e}{Combines data sets B12901 and U10635.  See discussion in text.}
\tablenotetext{f}{Includes data from observation S4057701, obtained in 2002.}
\end{deluxetable}


\begin{deluxetable}{llrrrrrcrrc} 
\tablecolumns{11} 
\tablewidth{0pt}
\tabletypesize{\scriptsize}
\tablecaption{\label{tab_old_detections} 3\sig\ Detections of \ion{O}{6} $\lambda 1032$ Emission near the Magellanic Clouds from Paper~II}
\tablehead{ 
\colhead{Sight}  & \colhead{Target} & \colhead{$l$}   & \colhead{$b$}   & \colhead{$t$\tablenotemark{b}} & & \colhead{Intensity\tablenotemark{d}} & \colhead{Wavelength\tablenotemark{e}} &
\colhead{FWHM\tablenotemark{f}}   & \colhead{$v_{\rm LSR}$} & \colhead{$E(B-V)$\tablenotemark{g}} \\
\colhead{Line}    & \colhead{Name\tablenotemark{a}} & \colhead{(deg)} & \colhead{(deg)} & \colhead{(s)} & \colhead{Bin\tablenotemark{c}} & \colhead{(kLU)}   & \colhead{(\AA)} &
\colhead{(km s$^{-1}$)} & \colhead{(km s$^{-1}$)}   & \colhead{(mag)} }
\startdata 
C07601 & NGC6752               & 336.5 & $-25.6$ & 56225 & \phn8 & $1.8 \pm 0.4$ & $1031.87 \pm 0.02$ & $  6 \pm 10$ & $ -17 \pm \phn6$ & 0.056 \\
C10503 & MS2318.2-4220         & 348.1 & $-66.3$ & 97954 & \phn8 & $1.8 \pm 0.4$ & $1031.92 \pm 0.05$ & $ 90 \pm 40$ & $  -5 \pm 13$ & 0.021 \\
D04001 & NGC625                & 273.7 & $-73.1$ & 43428 & \phn8 & $2.6 \pm 0.8$ & $1032.03 \pm 0.06$ & $ 90 \pm 50$ & $  19 \pm 18$ & 0.016 \\
D90305 & Fairall 917           & 346.6 & $-39.5$ & 21796 & 14 & $5.5 \pm 1.5$ & $1031.96 \pm 0.08$ & $180 \pm 50$ & $  11 \pm 24$ & 0.035 \\
I20505\tablenotemark{h} & DET-BKGND             & 278.6 & $-45.3$ & 46386 & 8 & $3.0 \pm 0.6$ & $1032.69 \pm 0.05$ & $100 \pm 20$ & $206 \pm 13$ & 0.160 \\
I20509 & Background            & 315.0 & $-41.3$ & 93437 & \phn8 & $3.3 \pm 0.6$ & $1032.05 \pm 0.05$ & $170 \pm 40$ & $  26 \pm 16$ & 0.036 \\
P10411 & WD0455-282            & 229.3 & $-36.2$ & 46240 & \phn8 & $4.3 \pm 0.9$ & $1031.95 \pm 0.03$ & $ 60 \pm 40$ & $ -12 \pm \phn9$ & 0.023 \\
P20422 & WD2004-605            & 336.6 & $-32.9$ & 76972 & \phn8 & $4.8 \pm 0.6$ & $1031.91 \pm 0.03$ & $130 \pm 30$ & $  -7 \pm \phn8$ & 0.049 \\
P20516 & LSE 44                & 313.4 & $ 13.5$ & 85811 & \phn8 & $3.5 \pm 0.6$ & $1031.83 \pm 0.04$ & $100 \pm 20$ & $ -28 \pm 12$ & 0.124 \\
S40506 & HD093840-BKGD         & 282.1 & $ 11.1$ & 12962 & \phn8 & $4.0 \pm 1.3$ & $1031.91 \pm 0.05$ & $ 20 \pm 100$ & $ -14 \pm 13$ & 0.167 \\
S50508 & LS982-BKGD            & 257.1 & $ -3.9$ & 10515 & \phn8 & $9.1 \pm 1.8$ & $1031.97 \pm 0.03$ & $ 30 \pm 30$ & $  -4 \pm \phn8$ & 1.260
\enddata 

\tablenotetext{a}{Target names are taken from the data file headers.}
\tablenotetext{b}{Exposure time is night only.}
\tablenotetext{c}{Spectral binning in 0.013 \AA\ pixels.}
\tablenotetext{d}{1\,LU = 1~photon~s$^{-1}$\,cm$^{-2}$\,sr$^{-1}$.}
\tablenotetext{e}{Wavelengths are heliocentric.}
\tablenotetext{f}{Gaussian FWHM values include the smoothing imparted by the instrument optics. Values less than $\sim$ 25 \kms\ indicate that the emission does not fill the LWRS aperture.}
\tablenotetext{g}{Extinction from \citealt{Schlegel:98}.}
\tablenotetext{h}{See Table 6 of \citetalias{Dixon:06}.}
\end{deluxetable}


\begin{deluxetable}{llrrrcc}
\tablecolumns{7}
\tablewidth{0pt}
\tablecaption{Upper Limits ($\leq 2.0$ kLU) on O VI $\lambda 1032$ Emission near the Magellanic Clouds from Paper~II\label{tab_old_limits}}
\tablehead{
\colhead{Sight} & \colhead{Target} & \colhead{$l$} & \colhead{$b$} & \colhead{$t$\tablenotemark{b}} & \colhead{$3\sigma$ Limit\tablenotemark{c}} & \colhead{$E(B-V)$\tablenotemark{d}} \\
\colhead{Line} & \colhead{Name\tablenotemark{a}} & \colhead{(deg)} & \colhead{(deg)} & \colhead{(s)} & \colhead{(kLU)} & \colhead{(mag)} }

\startdata
D90302 & IRAS03335-5626        & 269.4 &  $ -48.9 $  & 33946 &  1.5 &  0.025 \\
P10425 & HD128620/HD128621     & 315.7 &  $  -0.7 $  & 26651 &  1.6 &  6.505 \\
P20406 & WD2127-222            &  27.4 &  $ -43.8 $  & 18938 &  2.0 &  0.046 \\
S10102 & WD1634-573            & 329.9 &  $  -7.0 $  & 67118 &  1.2 &  0.342 \\
S40573 & HD35580-BKGD          & 264.2 &  $ -34.5 $  & 40611 &  1.3 &  0.040 \\
S40582 & CD-61\_1208-BKGD      & 270.1 &  $ -30.6 $  & 42039 &  1.2 &  0.052 \\
S52301 & WD2211-495-BKGD       & 345.8 &  $ -52.6 $  & 58547 &  1.2 &  0.015 \\
Z90702 & CTS0563               & 354.7 &  $ -49.9 $  & 15329 &  1.5 &  0.017 \\
Z90719 & ESO116-G18            & 276.2 &  $ -47.6 $  & 10485 &  2.0 &  0.075 \\
Z90737 & FAIRALL333            & 327.7 &  $ -23.8 $  & 29366 &  1.4 &  0.083
\enddata

\tablenotetext{a}{Target names are taken from the data file headers.}
\tablenotetext{b}{Exposure time is night only.}
\tablenotetext{c}{1\,LU = 1~photon~s$^{-1}$\,cm$^{-2}$\,sr$^{-1}$.}
\tablenotetext{d}{Extinction from \citealt{Schlegel:98}.}
\end{deluxetable}

\clearpage

\clearpage
{\plotone{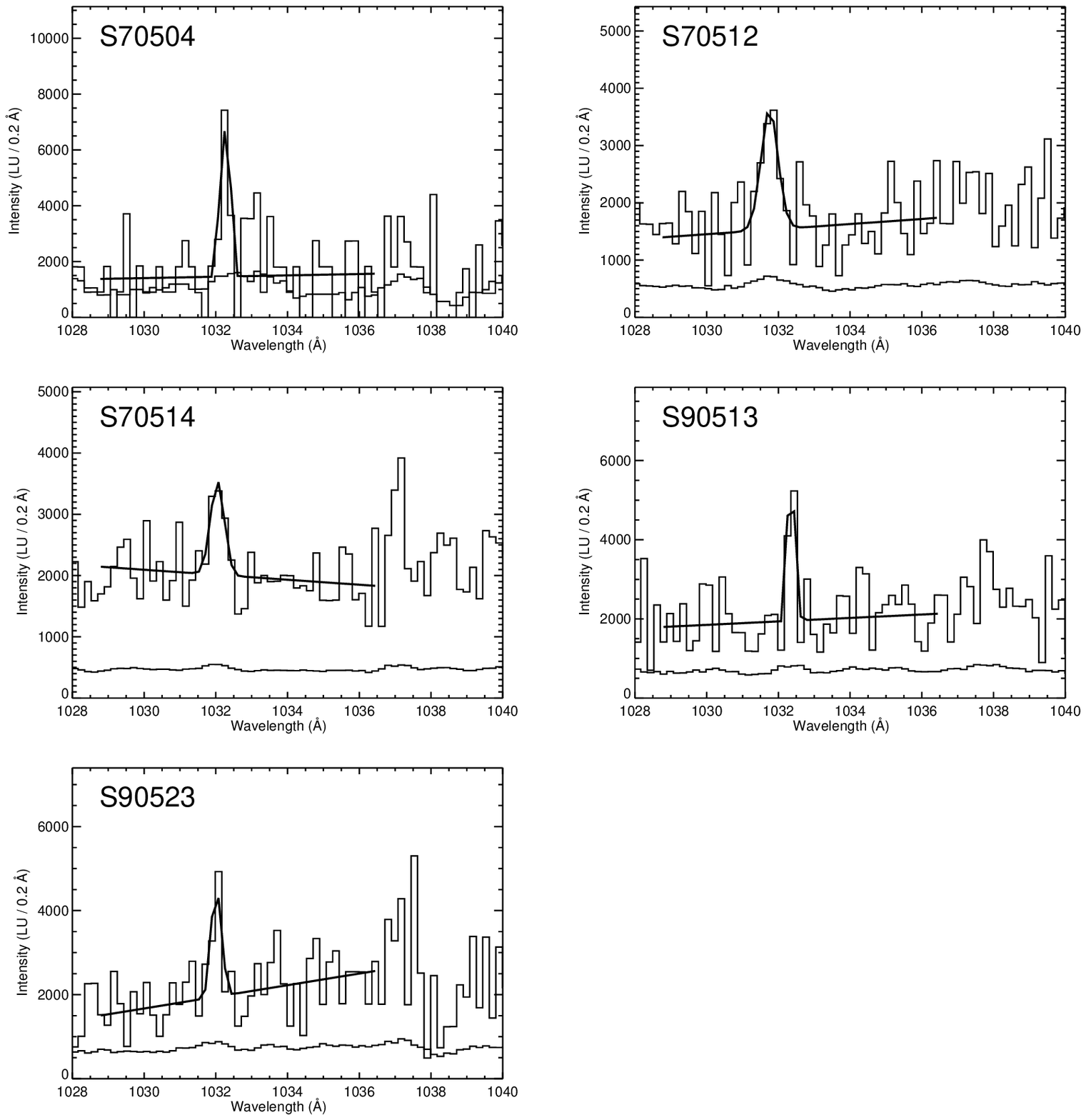}}\\
\centerline{{\it Fig. 1. --- Continued.}}

\clearpage
\clearpage

\epsscale{.90}
{\plotone{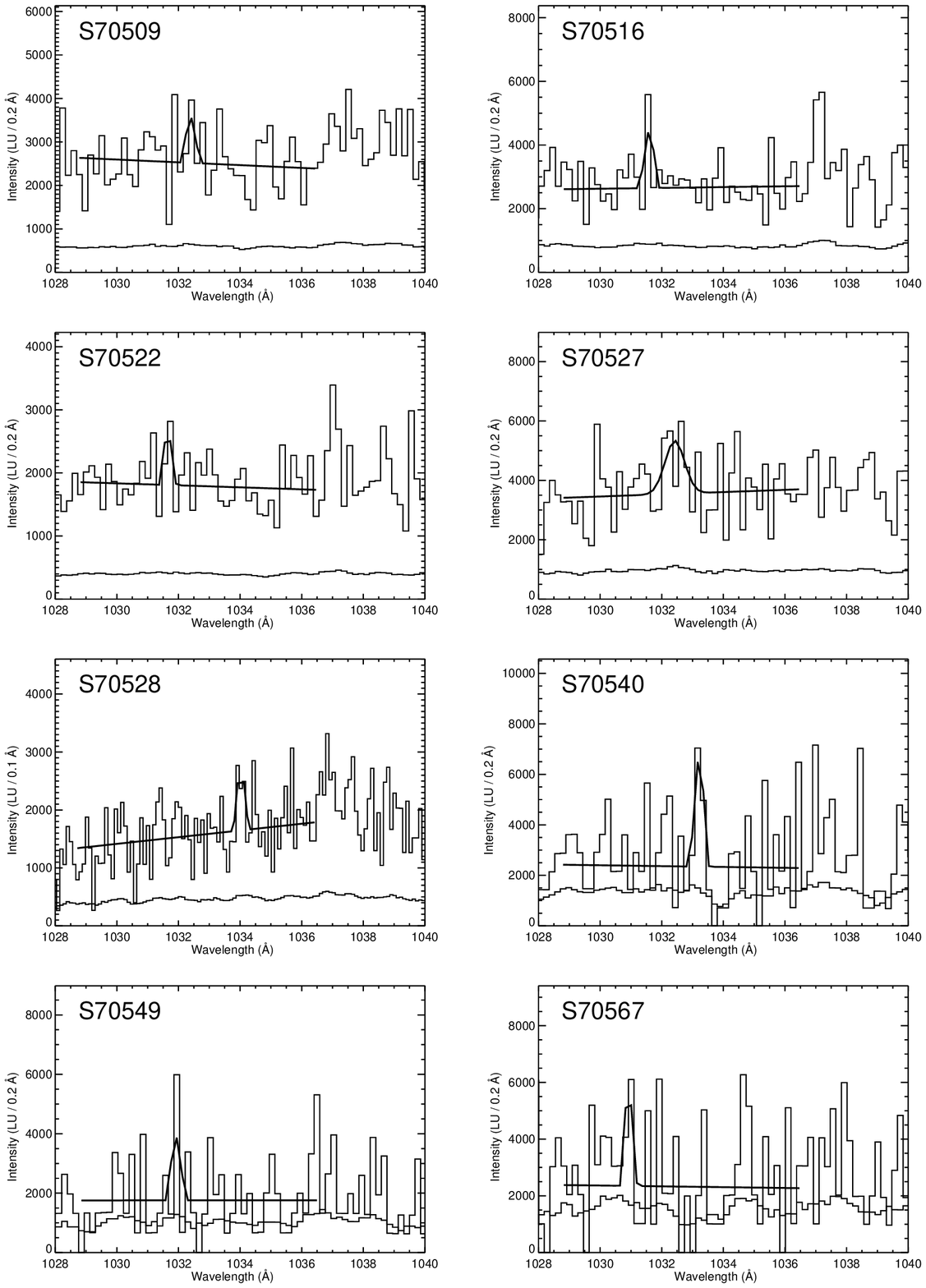}}\\
\centerline{{\it Fig. 2. --- Continued.}}
\clearpage

{\plotone{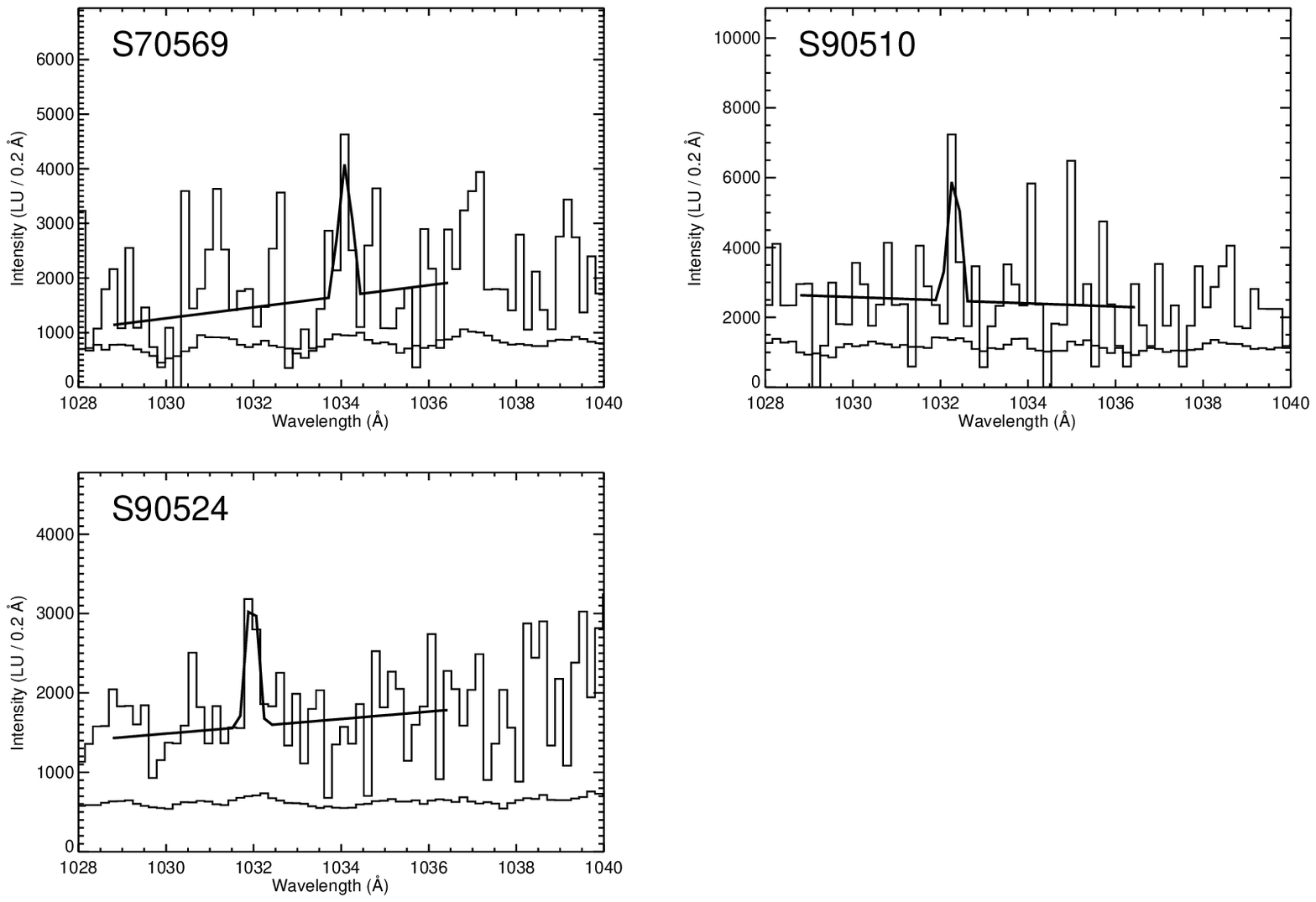}}\\
\centerline{{\it Fig. 2. --- Continued.}}
\clearpage

\clearpage

\clearpage

\end{document}